\pdfoutput=1
\RequirePackage{ifpdf}
\ifpdf % We are running pdfTeX in pdf mode
\documentclass[pdftex]{sigma}
\else
\documentclass{sigma}
\fi

\numberwithin{equation}{section}

\begin{document}

%\allowdisplaybreaks

\renewcommand{\thefootnote}{$\star$}

\renewcommand{\PaperNumber}{057}

\FirstPageHeading

\ShortArticleName{A 2+1-Dimensional Non-Isothermal Magnetogasdynamic System}

\ArticleName{A 2+1-Dimensional Non-Isothermal\\ Magnetogasdynamic
System.\\ Hamiltonian--Ermakov Integrable Reduction\footnote{This
paper is a contribution to the Special Issue ``Geometrical Methods in Mathematical Physics''. The full collection is available at \href{http://www.emis.de/journals/SIGMA/GMMP2012.html}{http://www.emis.de/journals/SIGMA/GMMP2012.html}}}

\Author{Hongli AN~$^\dag$ and Colin ROGERS~$^{\ddag\S}$}

\AuthorNameForHeading{H.L.~An and C.~Rogers}

\Address{$^\dag$~College of Science, Nanjing Agricultural University,
Nanjing 210095, P.R.~China}
\EmailD{\href{mailto:kaixinguoan@163.com}{kaixinguoan@163.com}}

\Address{$^\ddag$~School of Mathematics and Statistics, The
University of New South Wales,\\
\hphantom{$^\ddag$}~Sydney, NSW 2052, Australia}

 \Address{$^\S$~Australian Research Council Centre of Excellence
for Mathematics {\rm \&} Statistics\\
\hphantom{$^\S$}~of Complex Systems, School of
Mathematics, The University of New South Wales,\\
\hphantom{$^\S$}~Sydney, NSW2052, Australia}
\EmailD{\href{mailto:c.rogers@unsw.edu.au}{c.rogers@unsw.edu.au}}
\URLaddressD{\url{http://web.maths.unsw.edu.au/~colinr/}}

\ArticleDates{Received May 27, 2012, in f\/inal form August 02, 2012; Published online August 23, 2012}

\Abstract{A 2+1-dimensional anisentropic magnetogasdynamic system with a
polytropic gas law is shown to admit an integrable elliptic vortex
reduction when $\gamma= 2$ to a nonlinear dynamical subsystem with
underlying integrable Hamiltonian--Ermakov structure. Exact solutions
of the magnetogasdynamic system are thereby obtained which describe
a rotating elliptic plasma cylinder. The semi-axes of the elliptical
cross-section, remarkably, satisfy a~Ermakov--Ray--Reid system.}


\Classification{34A34; 35A25}

\section{Introduction}

Neukirch et al.~\cite{TN95,TNDC01,TNEP00} have
investigated 2+1-dimensional magnetogasdynamic systems via a
solution approach in which the nonlinear acceleration terms in the
Lundquist momentum equation either vanish or are conservative. By
contrast, in recent work~\cite{CR10,CRWS10} an elliptic vortex
ansatz was adopted in 2+1-dimensional isothermal magnetogasdynamics
and underlying integrable Ermakov--Ray--Reid structure was isolated.
In~\cite{CRWS10}, magnetogasdynamic pulsrodon-type solutions were
constructed analogous to those originally derived in elliptic
warm-core theory in~\cite{CR89}. The pulsrodons describe an
elliptical plasma cylinder bounded by a vacuum. The time-dependent
semi-axes of the elliptical cross-section of the cylinder were shown
to be governed by an integrable Hamiltonian Ermakov system.

The present work concerns an extension of that of~\cite{CR10} to a non-isothermal rotating magnetogasdynamic version
of a spinning non-conducting gas cloud system with origin in work of
Ovsiannikov~\cite{LO56} and Dyson~\cite{FD68}. A nonlinear dynamical
subsystem is derived which is again remarkably, shown to have
integrable Hamiltonian Ermakov--Ray--Reid structure. Moreover, a~Lax
pair for the dynamical system is constructed.

\section{The magnetogasdynamic system}

Here, we consider a 2+1-dimensional anisentropic magnetogasdynamic
system incorporating rotation, namely,
\begin{gather}  \frac{\partial \rho} {\partial t} +\operatorname{div}(\rho\mathbf{q}) =0 , \label{b1} \\
 \rho\left[ \frac{\partial \mathbf{q}} {\partial t} + (\mathbf{q}\cdot\nabla) \mathbf{q}+f \mathbf{k} \times
\mathbf{q}\right] - \mu\operatorname{curl} \mathbf{H} \times
\mathbf{H} + \nabla p=0 , \label{b2} \\
 \operatorname{div} \mathbf{H} = 0 , \label{b3} \\
 \frac{\partial \mathbf{H}}{\partial t} = \operatorname{curl} (\mathbf{q} \times \mathbf{H}), \label{b4} \\
 \frac{\partial S}{\partial t} + \mathbf{q}\cdot\nabla S = 0, \label{b5}
 \end{gather}
where the velocity $\mathbf{q}$ and magnetic f\/ield $\mathbf{H}$ are
given by
\begin{gather} \mathbf{q}=u\textbf{i}+v\textbf{j} , \qquad %\label{b6} \\
 \mathbf {H}=\nabla A\times \mathbf{k}+h \mathbf{k} \label{b7}
 \end{gather}
respectively, while the gas law adopts the polytropic form
\begin{gather}
S = -\ln \rho + \frac{1}{\gamma - 1} \ln T ,\qquad \gamma \neq 1  \label{b8}
\end{gather}
with
\begin{gather}
p =\rho T. \label{b9}
\end{gather}
In the above, the magneto-gas density $\rho(\mathbf{x},t)$, pressure
$p(\mathbf{x},t)$, entropy $S(\mathbf{x},t)$, tempera\-tu\-re~$T(\mathbf{x},t)$ and magnetic f\/lux $A(\mathbf{x},t)$ are all
 assumed to be dependent only on
 $\mathbf{x}=x\textbf{i}+y\textbf{j}$ and time~$t$. In addition, $f$ is the Coriolis constant, $\mu$ the magnetic permeability and~$h(\mathbf{x},t)$ the transverse component of the magnetic f\/ield.

  Insertion of the representation (\ref{b7}) into
Faraday's law (\ref{b4}) produces the convective constraint
\begin{gather}
\frac{\partial A}{\partial t} +\mathbf{q}\cdot\nabla A=0\label{b10}
\end{gather}
together with
\begin{gather*}
\frac{\partial h}{\partial t} +\operatorname{div}(h\mathbf{q})=0,%\label{b11}
\end{gather*}
which holds automatically if we set
\begin{gather*}
h=\lambda  \rho ,\qquad \lambda\in \mathbb{R} .%\label{b12}
\end{gather*}
Here, a novel two-parameter $(m, n)$ pressure-density
ansatz
\begin{gather}
p=\varepsilon_0(t)\rho^2+\varepsilon_1(t)\rho^n+\varepsilon_2(t)\rho^m.\label{b13}
\end{gather}
is introduced. In the magnetogasdynamic study of~\cite{TNDC01}, a
relation $p\sim\rho$ was adopted, while pressure-density relations
of the type $p\sim\rho^2$ arise in astrophysical contexts~\cite{Schafer}. A parabolic pressure-density law was recently
employed in 2+1-dimensional isothermal magnetogasdynamics in~\cite{CRWS10} and pulsrodon-type solutions were isolated.

In the present non-isothermal context, substitution
of (\ref{b13}) in (\ref{b9}) produces the temperature distribution
\begin{gather}\label{b14}
T=\varepsilon_0(t)\rho+\varepsilon_1(t)\rho^{n-1}+\varepsilon_2(t)\rho^{m-1},
\end{gather}
while the entropy distribution adopts the form
\begin{gather}
S=-\ln
\rho+\frac{1}{\gamma-1}\ln\big(\varepsilon_0(t)\rho+\varepsilon_1(t)\rho^{n-1}+\varepsilon_2(t)\rho^{m-1}\big).\label{b15}
\end{gather}

 The energy equation now requires that
\begin{gather*}
(\rho_t+\mathbf{q}\cdot\nabla\rho)\left[\frac{\varepsilon_0+(n-1)\varepsilon_1\rho^{n-2}+(m-1)\varepsilon_2\rho^{m-2}}{(\gamma-1)
(\varepsilon_0\rho+\varepsilon_1\rho^{n-1}+\varepsilon_2\rho^{m-1})}-\frac{1}{\rho}\right]\\
\qquad{}
+\frac{\dot{\varepsilon}_0\rho+\dot{\varepsilon}_1\rho^{n-1}+\dot{\varepsilon}_2\rho^{m-1}}{(\gamma-1)
(\varepsilon_0\rho+\varepsilon_1\rho^{n-1}+\varepsilon_2\rho^{m-1})}=0 ,
\end{gather*}
whence, on use of the continuity equation (\ref{b1})
\begin{gather}
-\operatorname{div}\mathbf{q}
\left[(2-\gamma)\varepsilon_0\rho+(n-\gamma)\varepsilon_1\rho^{n-1}\!+(m-\gamma)\varepsilon_2\rho^{m-1}\right]\!
+\dot{\varepsilon}_0\rho+\dot{\varepsilon}_1\rho^{n-1}\!+\dot{\varepsilon}_2\rho^{m-1}=0\!\!\!
\label{b16}
\end{gather}
that is
\begin{gather*}
\frac{1}{\gamma-1}\frac{\dot{T}}{T}
\left[(2-\gamma)\varepsilon_0\rho+(n-\gamma)\varepsilon_1\rho^{n-1}+(m-\gamma)\varepsilon_2\rho^{m-1}\right]
+\dot{\varepsilon}_0\rho+\dot{\varepsilon}_1\rho^{n-1}+\dot{\varepsilon}_2\rho^{m-1}=0 .%\label{b17}
\end{gather*}

  On substitution of (\ref{b7}) and (\ref{b13}) into
the momentum equation (\ref{b2}), it is seen that
\begin{gather*}
\frac{\partial\mathbf{q}} {\partial  t} + \mathbf{q}\cdot\nabla \mathbf{q}+f \mathbf{k} \times
\mathbf{q} +\frac{1}{\rho}\left[ \mu(\nabla^2A)(\nabla
A)+\varepsilon_1\nabla\rho^n\right]+(\mu\lambda^2+2\varepsilon_0)\nabla\rho+\frac{\varepsilon_2}{\rho} \nabla\rho^m=0
 %\label{b18}
\end{gather*}
together with
 \begin{gather*}
  A_y\rho_x-A_x\rho_y=0,%\label{b19}
\end{gather*}
so that
 \begin{gather*}
 A=A(\rho,t) .%\label{b20}
\end{gather*}
Attention is here restricted to the separable case
 \begin{gather*}
 A=\Phi(\rho)\Psi(t) %\label{b21}
\end{gather*}
whence, substitution into (\ref{b10}) and use of the continuity
equation yields
 \begin{gather*}
 \frac{\dot{\Psi}(t)}{\Psi(t)}=\rho\frac{\Phi'(\rho)}{\Phi(\rho)}\operatorname{div}\mathbf{q} .%\label{b22}
\end{gather*}
Here, we proceed with
\begin{gather*}
 \Phi=\rho^n, %\label{b23}
\end{gather*}
where $n$ is the parameter involving in the relation (\ref{b14}), so
that
\begin{gather}
 \operatorname{div}\mathbf{q}=\left(\frac{1}{n}\right)\frac{\dot{\Psi}}{\Psi} \label{b24}
\end{gather}
and
\begin{gather*}
 A=\rho^n\Psi(t) .%\label{b25}
\end{gather*}
Hence, as in the case of the spinning non-conducting gas cloud
analysis of Ovsiannikov~\cite{LO56} and Dyson~\cite{FD68}, the
divergence of the velocity is dependent only on time. Moreover, the
relation~(\ref{b16}) shows that
\begin{gather*}
-\frac{1}{n} \frac{\dot{\Psi}}{\Psi}
\left[(2-\gamma)\varepsilon_0\rho+(n-\gamma)\varepsilon_1\rho^{n-1}+(m-\gamma)\varepsilon_2\rho^{m-1}\right]
+\dot{\varepsilon}_0\rho+\dot{\varepsilon}_1\rho^{n-1}+\dot{\varepsilon}_2\rho^{m-1}=0
\end{gather*}
and it is observed that this condition holds identically with
\begin{gather}
\varepsilon_0=\alpha_0\Psi^{\frac{2-\gamma}{n}} ,\label{b26}
\\
\varepsilon_1=\alpha_1\Psi^{\frac{n-\gamma}{n}} ,\label{b27}
\\
\varepsilon_2=\alpha_2\Psi^{\frac{m-\gamma}{n}} ,\label{b28}
\end{gather}
where $\alpha_i$ $(i=0,1,2)$ are arbitrary constants of integration.
In addition, the isentropic condition~(\ref{b5})
 together with the polytropic gas law~(\ref{b8}) and the continuity
 equation~(\ref{b1}) show that
\begin{gather*}
 \operatorname{div}\mathbf{q}=\frac{1}{1-\gamma} \frac{\dot{T}}{T}
\end{gather*}
whence, on use of (\ref{b14}),
\begin{gather*}%\label{a}
 (2-\gamma)\operatorname{div}\mathbf{q}=\frac{\dot{\varepsilon}_0}{\varepsilon_0} ,\qquad
 (n-\gamma)\operatorname{div}\mathbf{q}=\frac{\dot{\varepsilon}_1}{\varepsilon_1} ,\qquad
 (m-\gamma)\operatorname{div}\mathbf{q}=\frac{\dot{\varepsilon}_2}{\varepsilon_2} .
\end{gather*}
It is seen that in view of (\ref{b24}), these relations are indeed
consistent with (\ref{b26})--(\ref{b28}).

  In summary, the magnetogasdynamic system now reduces
to consideration of the nonlinear coupled system
\begin{gather}
 \dfrac{\partial\rho}{\partial
 t}+\operatorname{div}(\rho\mathbf{q})=0 ,\nonumber
 \\
 \frac{\dot{\Psi}}{\Psi}=n\operatorname{div}\mathbf{q} , \label{b29}
\\
\dfrac{\partial \mathbf{q}} {\partial  t} + \mathbf{q}\cdot\nabla
\mathbf{q}+f \mathbf{k} \times \mathbf{q}
+\dfrac{1}{\rho}\left(\mu\Psi^2\nabla^2\rho^n+\varepsilon_1\right)\nabla\rho^n
+(\mu\lambda^2+2\varepsilon_0)\nabla\rho+\dfrac{\varepsilon_2m}{m-1}\nabla\rho^{m-1}=0, \nonumber
\end{gather}
where $m \neq 1$,
together with the additional conditions (\ref{b26})--(\ref{b28}). The
inherent nonlinearity of the system~(\ref{b29}) remains a major
impediment to analytic progress. It is noted also that the system (\ref{b29})$_{1,3}$ is overdetermined since it is implicitly constrained by
the requirement~(\ref{b29})$_2$ that $\operatorname{div}\mathbf{q}$ be a function of $t$ only.

\section{An elliptic vortex ansatz. A dynamical system reduction}

Here, integrable nonlinear dynamical subsystems of the
magnetogasdynamic system (\ref{b29}) are sought via an elliptic
vortex ansatz of the type
\begin{gather}
\mathbf{q} = \mathbf{L}(t)\mathbf{x} + \mathbf{M}(t) ,\qquad
\rho=
\big(\mathbf{x}^T \mathbf{E}(t) \mathbf{x}+\rho_0\big)^{m-1} ,\qquad m\neq1,
\qquad \mathbf{x} =  \begin{pmatrix} x - \bar{q}(t)
\\ y - \bar{p}(t)
\end{pmatrix} \label{b30}
\end{gather}
with
\begin{gather} {\bf L}(t) =   \begin{pmatrix} u_1(t) & u_2(t) \\  v_1(t) & v_2(t) \end{pmatrix},
\qquad {\bf E}(t) =   \begin{pmatrix} a(t) & b(t) \\  b(t) & c(t) \end{pmatrix},
\qquad {\bf M}(t) =  \begin{pmatrix} \dot{\bar{q}}(t) \\  \dot{\bar{p}}(t) \end{pmatrix} .\label{b31} \end{gather}

  Insertion of (\ref{b30}) into the continuity
equation yields
\begin{gather}
 \begin{pmatrix} \dot{a} \\  \dot{b} \\  \dot{c} \end{pmatrix} +  \begin{pmatrix}
 2u_1 + (m-1)(u_1+v_2)& 2v_1 & 0 \\  u_2 & m(u_1+v_2) & v_1 \\  0 & 2u_2 & 2v_2+(m-1)(u_1+v_2) \end{pmatrix}  \begin{pmatrix} a
 \\ b \\ c \end{pmatrix} = {\bf{0}} \label{b32}
 \end{gather}
together with
\begin{gather} \dot{\rho}_0 +\rho_0(m-1)(u_1 + v_2)=0 ,
 \label{b33}
 \end{gather}
whence
\begin{gather}
\rho_0 ={\rm const} \, \Psi^{(1-m)/n} .
 \label{b34}
 \end{gather}

  If we now proceed with
\begin{gather}
n=m-1  \label{a1}
\end{gather}
together with
\begin{gather}
 2\varepsilon_0 +\mu\lambda^2=0  \label{b35}
\end{gather}
and
\begin{gather}
\varepsilon_1 +2\mu\Psi^2(a+c)=0
 \label{b36}
 \end{gather}
then it is seen that (\ref{b29})$_3$ reduces to
\begin{gather}
\dfrac{\partial \mathbf{q}} {\partial  t} + \mathbf{q}\cdot\nabla
\mathbf{q}+f \mathbf{k} \times \mathbf{q}
+\dfrac{m}{m-1}\varepsilon_2\nabla\rho^{m-1}=0 .\label{b37}
\end{gather}
The relation (\ref{b35}) implies that $\dot{\varepsilon}_0= 0$
whence~(\ref{b26}) shows that the adiabatic index $\gamma = 2$,
while~(\ref{b36}) and (\ref{b27}) together require
\begin{gather*}
a+c=-\frac{\alpha_1}{2\mu}\Psi^{-(n+2)/n}=-\frac{\alpha_1}{2\mu}\Psi^{(1+m)/(1-m)} ,\qquad n\neq1 .%\label{b38}
\end{gather*}

Substitution of (\ref{b30}) into (\ref{b37}) now gives
\begin{gather}
 \begin{pmatrix} \dot{u}_1 \\  \dot{u}_2 \\  \dot{v}_1 \\  \dot{v}_2 \end{pmatrix}
 +  \begin{pmatrix} {\bf{L}}^T & -f{\bf{I}} \\  f{\bf{I}} & {\bf{L}}^T \end{pmatrix}  \begin{pmatrix} u_1 \\  u_2
  \\  v_1 \\  v_2 \end{pmatrix} + 2 \varepsilon_2\frac{ m}{m-1} \begin{pmatrix} a \\  b \\  b \\  c \end{pmatrix} = \bf{0} \label{b39}
\end{gather}
augmented by the auxiliary linear equations
\begin{gather} \ddot{\bar{p}} + f \dot{
\bar{q}} = 0 ,\qquad \ddot{\bar{q}} - f \dot{\bar{p}} = 0 .
\label{b40}
\end{gather}
It is noted that the relation (\ref{b29})$_2$ together with
(\ref{a1}) shows that
\begin{gather*}\dot{\Psi} = (m-1)(u_1+v_2)\Psi . %\label{a2}
\end{gather*}
While $\rho_0$ is given in terms of $\Psi$ via (\ref{b34}). The
constraints (\ref{b35}) and (\ref{b36}) are to be adjoined and their
admissibility will be examined subsequently.

  In what follows, it proves convenient to proceed in
terms of new variables, namely
\begin{gather*}  G =u_1 + v_2 , \qquad G_R = \frac{1}{2} (v_1 - u_2) ,  \qquad
 G_S =\frac{1}{2} (v_1 + u_2) , \qquad G_N = \frac{1}{2} (u_1 - v_2) ,\\ %\label{b42}\\
 B= a + c ,\qquad B_S = b ,\qquad B_N = \frac{1}{2} (a-c) .
\end{gather*}
These quantities were originally introduced in a hydrodynamic
context (see e.g.~\cite{CR89}). Therein, $G$~and~$G_R$ correspond,
in turn, to the divergence and spin of the velocity f\/ield, while~$G_S$~and~$G_N$ represent shear and normal deformation rates. The
system~(\ref{b32}) and (\ref{b33}) together with (\ref{b39}) now reduces
to the nonlinear dynamical system
\begin{gather}
\begin{split}
& \dot{\rho}_0 + (m-1)\rho_0 G = 0 , \\
& \dot{B} + mBG + 4 ( B_N G_N + B_S G_S ) = 0, \\
& \dot{B}_S + mB_S G +  BG_S - 2B_N G_R = 0, \\
& \dot{B}_N + mB_N G + BG_N  + 2B_S G_R = 0, \\
& \dot{G} + \dfrac{1}{2}G^2 + 2 \big( G^2_N + G^2_S - G^2_R \big) - 2f G_R + 2\dfrac{\varepsilon_2m}{m-1}B = 0, \\
& \dot{G}_N + G G_N - f G_S + 2\dfrac{\varepsilon_2m}{m-1}B_N = 0 , \\
& \dot{G}_S + G G_S + f G_N + 2\dfrac{\varepsilon_2m}{m-1}B_S = 0 , \\
& \dot{G}_R + G G_R +\dfrac{1}{2} fG = 0
\end{split}   \label{b43}
\end{gather}
together with
\begin{gather}
\dot{\Psi} = (m-1)\Psi G. \label{b44}
 \end{gather}
It is observed that the introduction of the pressure-density
parameters $(m,n)$ and $\varepsilon_2$ leads to a~generalisation of
the nonlinear dynamical systems obtained in~\cite{CR10,CR89,CRHA10,CRWS10}.

  If we now introduce the quantity $\Omega$ via
\begin{gather*}
G = \frac{2 \dot{\Omega}}{\Omega} %\label{b45}
\end{gather*}
then (\ref{b43})$_1$ and (\ref{b43})$_8$ show, in turn, that
\begin{gather}
\rho_0 = \frac{c_\mathrm{I}}{\Omega^{2(m-1)}} \label{b46}
\end{gather}
and
\begin{gather}
G_R= \frac{c_0}{\Omega^2}- \frac{1}{2}f. \label{b47}
\end{gather}
While the relation (\ref{b44}) yields
\begin{gather} \Psi = \nu \Omega^{2(m-1)}, \label{b48}
\end{gather}
where $c_0$, $c_{\mathrm{I}}$ and $\nu$ denote arbitrary constants of
integration.

  Two conditions which are key to the subsequent
development and which may be established by appeal to the original
system~(\ref{b43}) are now recorded. These represent extensions of
results obtained in a hydrodynamic context~\cite{CR89,CRHA10}.

\begin{theorem}\label{theoremI}
\begin{gather*}
   \dot{M}+(m+1)GM=0,\qquad
\dot{Q}+(m+1)GQ =0, \label{5.3.1}
\end{gather*}
where
\begin{gather*}
 M = 2(B_NG_S-B_SG_N)-B\left(G_R+\dfrac{1}{2}f\right) , \qquad \triangle
=\dfrac{1}{4}B^2-B^2_S-B^2_N ,\\
 Q = -B\left(G^2_S+G^2_N+G^2_R+\dfrac{1}{4}G^2\right)+4G_R(B_NG_S-B_SG_N)\\
\hphantom{Q=}{} +2G(B_SG_S+B_NG_N) + 4
\dfrac{\varepsilon_2m}{m-1}\triangle
-4\dfrac{m}{m-1}\triangle\Omega^{m-1}  \int
\dot{\varepsilon}_2\Omega^{1-m} dt .
 %\label{5.3.1a}
 \end{gather*}
\end{theorem}

 New $\Omega$-modulated variables involving the
pressure-density parameter~$m$ are now introduced according to
\begin{gather}   \bar{B} = \Omega^{2m} B ,\!\!\qquad \bar{B}_S = \Omega^{2m} B_S ,\!\!\qquad \bar{B}_N = \Omega^{2m} B_N , \!\!\qquad
\bar{G}_S = \Omega^2 G_S ,\!\!\qquad \bar{G}_N = \Omega^2 G_N,\!\!\!
\label{5.3.9}
\end{gather}
whence the dynamical system (\ref{b43}) reduces to
\begin{gather}
\begin{split}
&  \dot{\bar{B}} + \dfrac{4 (\bar{B}_N \bar{G}_N + \bar{B}_S \bar{G}_S)}{\Omega^2}= 0 , \\
&  \dot{\bar{B}}_S + f \bar{B}_N+\dfrac{ \bar{B} \bar{G}_S - 2c_0 \bar{B}_N}{\Omega^2}= 0 , \\
&  \dot{\bar{B}}_N - f \bar{B}_S+ \dfrac{\bar{B} \bar{G}_N + 2c_0 \bar{B}_S}{ \Omega^2} = 0 , \\
&  \dot{\bar{G}}_S + f \bar{G}_N + \dfrac{2\varepsilon_2m}{m-1}\dfrac{\bar{B}_S}{ \Omega^{2(m-1)}} = 0 , \\
&  \dot{\bar{G}}_N - f \bar{G}_S +
\dfrac{2\varepsilon_2m}{m-1}\dfrac{\bar{B}_N} {\Omega^{2(m-1)}} = 0
\end{split} \label{5.3.10}
\end{gather}
augmented by the relations (\ref{b46}) and (\ref{b47}) together with
a nonlinear equation for $\Omega$, namely
\begin{gather}
\Omega^3 \ddot{\Omega} + \dfrac{1}{4} f^2\Omega^4 +\bar{G}^2_N +
\bar{G}^2_S- c^2_0 +\dfrac{\varepsilon_2m}{m-1}\dfrac{\bar{B}}{
\Omega^{2(m-2)}} = 0 .\label{5.3.11}
\end{gather}
 The reduced dynamical system (\ref{5.3.10}) together
with (\ref{5.3.11}) and the constraints given by
(\ref{b26})--(\ref{b28}) and (\ref{b35}), (\ref{b36}) will now be
examined in detail.

  Thus, if we turn to the expressions for
$\varepsilon_0$, $\varepsilon_1$ and $\varepsilon_2$ as given by
(\ref{b26})--(\ref{b28}), it is seen immediately that consistency of
(\ref{b26}) and (\ref{b35}) requires that the adiabatic index
$\gamma=2$. Further, comparison of the expressions for
$\varepsilon_1$ in~(\ref{b27}) and~(\ref{b36}) now yields
\begin{gather*}
\alpha_1\Psi^\frac{n-2}{n}+2\mu\Psi^2(a+c)=0,
\end{gather*}
where  the relations (\ref{a1}), (\ref{b48}) and (\ref{5.3.9})
combine to show that
\begin{gather*}
\alpha_1\nu^\frac{m-3}{m-1}+2\mu\nu^2\Omega^2\bar{B}=0, %\label{a52}
\end{gather*}
whence
\begin{gather*}
\nu=0\qquad {\rm or}\qquad \Omega^2\bar{B}=-\frac{\alpha_1}{2\mu}\nu^\frac{1+m}{1-m}:=\delta .%\label{a53}
\end{gather*}
In the former case, by virtue of (\ref{b48}), the magnetic f\/lux
$A$ vanishes so that the magnetic f\/ield~$\bf{H}$ is purely
transverse and the dynamical system (\ref{5.3.10}) and (\ref{5.3.11}) is
not thereby constrained. Here, we proceed with the latter case, so
that the system (\ref{5.3.10}) and (\ref{5.3.11}) is additionally
constrained by the requirement $\Omega^2 \bar{B}={\rm const}$ and
(\ref{5.3.10})$_1$ yields
\begin{gather*}
\bar{B}_N \bar{G}_N + \bar{B}_S
\bar{G}_S-\frac{\delta\dot{\Omega}}{2\Omega}=0 .%\label{a54}
\end{gather*}
Finally, for $\varepsilon_2$, the relations (\ref{b28}), (\ref{a1})
and (\ref{b48}) combine to show that
\begin{gather*}
\varepsilon_2=\alpha_2\nu^\frac{m-2}{m-1}\Omega^{2(m-2)},
\end{gather*}
which it subsequently proves convenient to re-write in this form
\begin{gather}
\varepsilon_2 = \alpha \left(\frac{m-1}{m} \right)
\Omega^{2(m-2)} .\label{a55}
\end{gather}

\section{Integrals of motion and parametrisation}

Under the constraint (\ref{a55}), the nonlinear dynamical system
(\ref{5.3.10}) is readily shown to admit the key integrals of motion
\begin{gather} \bar{B}^2_S + \bar{B}^2_N - \frac{1}{4}\bar{B}^2 = c_{\mathrm{II}} , \label{5.3.14} \\
 \bar{G}^2_S + \bar{G}^2_N - \alpha\bar{B} = c_{\mathrm{III}} , \label{5.3.15} \\
 2(\bar{B}_N \bar{G}_S - \bar{B}_S\bar{G}_N)-c_0\bar{B}=c_{\mathrm{IV}} ,\label{a56}\\
 2(G_R+c_0\bar{B})+2G(\bar{B}_S\bar{G}_S+\bar{B}_N\bar{G}_N)+4\alpha c_{\mathrm{II}}\Omega^{-2}\frac{m-1}{m-3}\nonumber\\
 \qquad{}
-\bar{B}\Omega^2\left(\frac{\bar{G}^2_S + \bar{G}^2_N
}{\Omega^4}+\frac{G^2}{4}+G^2_R\right)=c_{\mathrm{V}} ,\label{a57}
\end{gather}
where $c_{\mathrm{II}}$, $c_{\mathrm{III}}$, $c_{\mathrm{IV}}$ and
$c_{\mathrm{V}}$ are constants of integration.

  The relations (\ref{5.3.14}) and (\ref{5.3.15}) may
be conveniently parametrised according to
\begin{alignat}{3}
 & \bar{B}_S = - \sqrt{c_{\mathrm{II}} + \dfrac{1}{4} \bar{B}^2}  \cos \phi(t) , \qquad & & \bar{B}_N = - \sqrt{c_{\mathrm{II}} + \dfrac{1}{4} \bar{B}^2}  \sin \phi(t) , & \nonumber\\
& \bar{G}_S = - \sqrt{c_{\mathrm{III}} + \bar{B}}  \sin \theta(t) ,\qquad && \bar{G}_N = + \sqrt{c_{\mathrm{III}} + \bar{B}}  \cos
\theta(t). &
  \label{5.4.1}
\end{alignat}
Substitution of this parametrisation into (\ref{5.3.10})$_1$ yields
\begin{gather} \dot{\bar{B}} + \frac{4}{\Omega^2} \sqrt{ ( c_{\mathrm{II}} + \bar{B}^2 / 4  )(  c_{\mathrm{III}} + \alpha\bar{B}  )} \sin(\theta - \phi) = 0, \label{5.4.4}
\end{gather}
while conditions (\ref{5.3.10})$_{2,3}$ reduce to a single relation,
namely
\begin{gather}
\left(\dot{\phi} - f+ \frac{2c_0}{\Omega^2}\right)\sqrt{c_{\mathrm{II}} + \frac{1}{4}\bar{B}^2} - \frac{\bar{B}}{\Omega^2} \sqrt{c_{\mathrm{III}} + \alpha\bar{B}} \cos(\theta - \phi) = 0. \label{5.4.5}
\end{gather}
Similarly, (\ref{5.3.10})$_{4,5}$ produce another single requirement
\begin{gather}
 (\dot{\theta}-f )\sqrt{c_{\mathrm{III}} + \alpha\bar{B}} +\frac{2\alpha}{\Omega^2} \sqrt{c_{\mathrm{II}} + \frac{1}{4}\bar{B}^2} \cos(\theta - \phi) = 0 . \label{5.4.6}
  \end{gather}
 Moreover, substitution of the representations
(\ref{5.4.1}) into (\ref{a56}) yields
\begin{gather}
 c_0 \bar{B} = -c_{\mathrm{IV}} + 2\sqrt{\left(c_{\mathrm{II}} + \bar{B}^2/4 \right) \left(c_{\mathrm{III}} + \alpha\bar{B} \right)} \cos(\theta - \phi) , \label{5.4.8}
\end{gather}
while  elimination of $\theta - \phi$ in (\ref{5.4.5}) and
(\ref{5.4.6}) respectively shows that
\begin{gather}
 \dot{\phi} = f + \frac{2}{\Omega^2}\left[\frac{\delta(c_0\delta + c_{\mathrm{IV}}\Omega^2)}{\delta^2+4c_{\mathrm{II}}\Omega^4}-c_0\right]
 \label{5.4.9}
\end{gather}
and
\begin{gather}
\dot{\theta} = f - \frac{\alpha}{\Omega^2}\left[\frac{c_0\delta+
c_{\mathrm{IV}}\Omega^2}{\alpha\delta^2+c_{\mathrm{III}}\Omega^2}\right].
\label{5.4.10}
\end{gather}

It remains to consider the nonlinear equation
(\ref{5.3.11}) for $\Omega$, namely
\begin{gather*}
\Omega^3\ddot{\Omega}+\frac{f^2}{4}\Omega^4+c_\mathrm{{III}}-c^2_0
+\dfrac{\varepsilon_2m}{m-1}\dfrac{\bar{B}}{ \Omega^{2(m-2)}}=0
\end{gather*}
which, by virtue of (\ref{a55}), reduces to a generalisation of the
classical Steen--Ermakov equa\-tion~\cite{VE80,Steen}, namely
\begin{gather}
 \ddot{\Omega} + \dfrac{1}{4} f^2\Omega=\frac{c^2_0-c_\mathrm{{III}}}{\Omega^3}-\frac{2\alpha\delta}{\Omega^5} .\label{5.3.20}
\end{gather}
Further, on use of Theorem~\ref{theoremI}, it may be readily shown that
there is the necessary requirement (cf.~\cite{CR89})
\begin{gather*}
\ddot{(\Omega^2\overline{B})} + f^2 \Omega^2 \bar{B} = -2(Q^* +
fM^*)\Omega^{2(m+1)} = -2(c_{\rm V} + f c_{\mathrm{IV}}) . %\label{5.4.11}
\end{gather*}
This holds automatically here with
\begin{gather}
\Omega^2 \bar{B}= {\rm const} =\delta = -2(c_{\mathrm{V}} + fc_{\mathrm{IV}}) / f^2,\qquad f\neq0 . \label{5.4.12}
\end{gather}

  Elimination of $\theta - \phi$ between (\ref{5.4.4})
and (\ref{5.4.8}) now yields
\begin{gather*}
\dot{\bar{B}}^2+\frac{4}{\Omega^4}(c_0\bar{B}+c_{\mathrm{IV}})^2=\frac{4}{\Omega^4}\big(4c_{\mathrm{II}}+\bar{B}^2\big)(c_{\mathrm{III}}+\alpha\bar{B}),
%\label{5.4.13}
\end{gather*}
whence on use of (\ref{5.4.12})
\begin{gather}
\delta^2\dot{\Omega}^2+\frac{(c_0^2-c_{\mathrm{III}})\delta^2}{\Omega^2}+\big(c^2_{\mathrm{IV}}
-4c_\mathrm{{II}}c_\mathrm{{\mathrm{III}}}\big)\Omega^2-\frac{\alpha\delta^3}{\Omega^4}+2c_0c_\mathrm{{IV}}-4\alpha
c_{\mathrm{II}}\delta=0 .\label{5.4.15}
\end{gather}
The latter equation is required to be compatible with the $1^{\rm st}$
integral of (\ref{5.3.20}), namely
\begin{gather*}
\dot{\Omega}^2+\frac{1}{4}f^2\Omega^2+\frac{(c_0^2-c_{\mathrm{III}})}{\Omega^2}-\frac{\alpha\delta}{\Omega^4}+k=0 %\label{5.4.16}
\end{gather*}
and these are indeed seen to be consistent subject to the relations
\begin{gather*}
c^2_\mathrm{{IV}}-4c_\mathrm{{II}}c_\mathrm{{III}}=\frac{\delta^2f^2}{4} ,%\label{5.4.17}
\end{gather*}
and
\begin{gather*}
k=\frac{2c_0c_\mathrm{{IV}}-4\alpha
c_\mathrm{{II}}\delta}{\delta^2} . %\label{5.4.17a}
\end{gather*}

  In summary, a multi-parameter class of exact vortex
solutions of the original
 2+1-dimensional magnetogasdynamic system has been generated with the velocity components $u_1$, $u_2$, $v_1$, $v_2$ and the
quantities $a$, $b$, $c$, $\rho_0$ in the density representation given, in
turn, by
\begin{gather}
 u_1 = \dfrac{\dot{\Omega}}{\Omega} + \dfrac{1}{\Omega^3} \sqrt{\alpha\delta+c_{\mathrm{III}}\Omega^2} \cos\theta(t) ,
 \qquad  v_1 =\dfrac{c_0}{\Omega^2}- \dfrac{f}{2}
 -\dfrac{1}{\Omega^3} \sqrt{\alpha\delta+c_{\mathrm{III}}\Omega^2} \sin\theta(t), \nonumber\\
u_2 = - \dfrac{c_0}{\Omega^2} + \dfrac{f}{2}-\dfrac{1}{\Omega^3}
\sqrt{\alpha\delta+c_{\mathrm{III}}\Omega^2} \sin\theta(t)
 ,\qquad v_2 = \dfrac{\dot{\Omega}}{\Omega} -
\dfrac{1}{\Omega^2}
\sqrt{\alpha\delta+c_{\mathrm{III}}\Omega^2} \cos\theta(t)
  \label{5.4.18}
\end{gather}
together with
\begin{gather*}   a = \dfrac{1}{2\Omega^{2(m+1)}} \left[\delta- \sqrt{4c_{\mathrm{II}}\Omega^4 + \delta^2} \sin\phi(t)\right] ,\qquad
b = \dfrac{1}{2\Omega^{2(m+1)}}\sqrt{4c_{\mathrm{II}}\Omega^4 +
\delta^2} \cos\phi(t) ,\\
c = \dfrac{1}{\Omega^{2(m+1)}} \left[\delta+\sqrt{4c_{\mathrm{II}}\Omega^4 + \delta^2} \sin\phi(t)\right] ,\qquad
\rho_0 = \dfrac{c_{\mathrm{I}}}{\Omega^{2(m-1)}}, %\label{5.4.19}
\end{gather*}
 where the angles $\phi$ and $\theta$ are obtained by integration of (\ref{5.4.9}) and (\ref{5.4.10}), respectively while $\Omega$ is given by an elliptic integral resulting from (\ref{5.4.15}).

  The magnetic f\/lux $A$ is given by
 \begin{gather*}
 A=\nu\rho^{m-1}\Omega^{2(m-1)}=\nu\left[a(x-\bar{q})^2+2b(x-\bar{q})(y-\bar{p})+c(y-\bar{p})^2+\rho_0\right]\Omega^{2(m-1)}, %\label{5.4.21}
\end{gather*}
while the temperature $T$ and entropy distribution $S$ are determined by~(\ref{b14}) and~(\ref{b15}), respectively.

\section{Hamiltonian Ermakov structure}

 The nonlinear dynamical system (\ref{b43}) may be shown to have remarkable underlying
 structure in that it will be seen to reduce to consideration of an integrable Ermakov--Ray--Reid type system
\begin{gather*} \ddot{\alpha} + \omega^2(t)\alpha =
\dfrac{1}{\alpha^2\beta}F(\beta/\alpha) ,\qquad
\ddot{\beta} + \omega^2(t)\beta =
\dfrac{1}{\alpha\beta^2}G(\alpha/\beta) .  %\label{d1}
\end{gather*}
Such systems have their origin in the work of Ermakov~\cite{VE80}
and were introduced by Ray and Reid in~\cite{JR80,JRJR80}. Extension
to 2+1-dimensions were presented in~\cite{CRCHJR93} and to
multi-component systems in~\cite{CRWS96}. The main theoretical
interest in the system resides in its admittance of a distinctive
integral of motion, namely, the Ray--Reid invariant
\begin{gather*}
I = \dfrac{1}{2}(\alpha\dot{\beta} - \beta\dot{\alpha})^2 + \int^{\beta/\alpha}F(z)dz
 + \int^{\alpha/\beta}G(w)dw \label{d2} .
 \end{gather*}
Ermakov--Ray--Reid systems arise, in particular, in a variety of
contexts in nonlinear optics (see e.g.~\cite{JPA12,JPA} and
references cited therein).

  Here, we proceed with $\bar{p}(t) = \bar{q}(t) = 0$
in the ansatz~(\ref{b30}), since the translation terms~$\bar{p}(t)$,~$\bar{q}(t)$ are readily re-introduced by use of a Lie group
invariance of the magnetogasdynamic system.

  The semi-axes of the time-modulated ellipse
\begin{gather*}   a(t)x^2 + 2b(t)xy + cy^2 + h_0(t) = 0, \qquad
 ac - b^2 > 0,  %\label{5.5.3}
 \end{gather*}
are now given by
\begin{gather*}
\Phi = \sqrt{\frac{2\rho_0}{\sqrt{(a-c)^2 + 4b^2} - (a + c)}}= \sqrt{-\frac{c_{\mathrm{I}}}{2c_{\mathrm{II}}}}\sqrt{-\delta-\sqrt{4c_{\mathrm{II}}\Omega^4+\delta^2}}%\label{5.5.4}
\end{gather*}
and
\begin{gather*}
\Psi = \sqrt{\frac{2\rho_0}{-\sqrt{(a-c)^2 + 4b^2} - (a + c)}}= \sqrt{-\frac{c_{\mathrm{I}}}{2c_{\mathrm{II}}}}\sqrt{-\delta+\sqrt{4c_{\mathrm{II}}\Omega^4+\delta^2}},
%\label{5.5.42}
\end{gather*}
where it is required that
\begin{gather*}
c_{\mathrm{I}}>0,\qquad c_{\mathrm{II}}<0,\qquad \delta<0,\qquad \delta^2+4c_{\mathrm{II}}\Omega^4>0 .%\label{5.5.53}
\end{gather*}

  It is readily established that the semi-axes
$\Phi$, $\Psi$ are governed by a Ermakov--Ray--Reid system, namely
\begin{gather}
 \ddot{\Phi} + \dfrac{1}{4} f^2 \Phi = \dfrac{1}{\Phi^2\Psi} \left[  \dfrac{ZZ'}{1+( \Psi/\Phi )^2} - \left(\dfrac{\Psi}{\Phi} \right) \dfrac{(Z^2 + k/4 )}{[ 1+( \Psi/\Phi )^2 ]^2}  \right],\nonumber \\
\ddot{\Psi} + \dfrac{1}{4} f^2 \Psi = \dfrac{1}{\Phi\Psi^2} \left[
- \dfrac{ZZ'}{1+( \Psi/\Phi )^2}-\left( \dfrac{\Phi}{\Psi}
 \right) \dfrac{(Z^2 + k/4 )}{[ 1+(\Phi/\Psi)^2 ]^2} \right],
  \label{5.5.5}
\end{gather}
where
\begin{gather*}
Z=Z(\Phi/\Psi)=\Psi \dot{\Phi} - \dot{\Psi}\Phi = \frac{2c_{\mathrm{I}}}{\Omega\sqrt{-c_{\mathrm{II}}}}
\sqrt{\frac{(\delta^2 + 4c_{\mathrm{II}}\Omega^4)(\alpha\delta +
c_{\mathrm{III}}\Omega^2) - \Omega^2(c_0\delta +
c_{\mathrm{IV}}\Omega^2)^2}{\delta^2 + 4c_{\mathrm{II}}\Omega^4}}
%\label{5.5.6}
\end{gather*}
and $\Omega$ is given in terms of the ratio of the semi-axes via the
relation
\begin{gather*}
\Omega = \left(-\frac{\delta^2}{c_{\mathrm{II}}}\right)^{1/4}\left(\frac{\Psi}{\Phi} + \frac{\Phi}{\Psi}\right)^{-1/2} . %\label{5.5.7}
\end{gather*}
In addition, the Ermakov--Ray--Reid system (\ref{5.5.5}) is seen to be
Hamiltonian with invariant
\begin{gather*} H = \frac{1}{2}\big(\dot{\Phi}^2 + \dot{\Psi}^2\big) - \frac{1}{2(\Phi^2 + \Psi^2)}\left[ Z^2 - \frac{f^2}{4} \big(\Phi^2 + \Psi^2\big)^2
 + \frac{k}{4}\right]=-\frac{1}{4}f^2\frac{c_{\mathrm{I}}c_{\mathrm{Iv}}}{c_{\mathrm{II}}} ,\qquad  c_{\mathrm{II}}\neq0,
 %\label{5.5.9}
 \end{gather*}
and accordingly, is amenable to the general procedure described in
detail in~\cite{CRHA10}.

  It is remarkable indeed that the semi-axes $\Phi$
and $\Psi$ of the time modulated ellipse associated with the density
representation in~(\ref{b30}), are governed by an integrable
Ermakov--Ray--Reid system, albeit of some complexity. In fact, a
Ermakov--Ray--Reid system may also be associated with the velocity
components, at least, in a particular reduction. Attention is here
restricted, as in the work of Dyson~\cite{FD68} on non-conducting
gas clouds, to irrotational motions in the absence of a Coriolis
term.

  Thus, here we set
\begin{gather*}
{\bf L} =  \begin{pmatrix} \dot{\alpha}(t)/\alpha(t)  & 0 \\  0 &  \dot{\beta}(t)/\beta(t) \end{pmatrix},
\qquad {\bf E} =  \begin{pmatrix} a(t) & 0 \\  0 & c(t) \end{pmatrix} %\label{5.7.1}
\end{gather*}
in (\ref{b31}) corresponding to the subclass of exact solutions in
(\ref{5.4.18}) with $\theta=0$, $\phi=\pi/2$ and
\begin{gather*} \dfrac{\dot{\alpha}}{\alpha} = \dfrac{\dot\Omega}{\Omega} + \dfrac{1}{\Omega^2}\sqrt{c_\mathrm{{III}} + \dfrac{\alpha\delta}{\Omega^2}} ,
\qquad \dfrac{\dot{\beta}}{\beta} = \dfrac{\dot\Omega}{\Omega} - \dfrac{1}{\Omega^2}\sqrt{c_\mathrm{{III}} + \dfrac{\alpha\delta}{\Omega^2}} , \\
a = \dfrac{1}{2\Omega^{2(m+1)}}\left[\delta -
\sqrt{4c_\mathrm{{II}}\Omega^4 +\delta^2}\right] ,\qquad c =
\dfrac{1}{2\Omega^{2(m+1)}}\left[\delta +
\sqrt{4c_\mathrm{{II}}\Omega^4 +\delta^2}\right] .
  %\label{5.7.2}
\end{gather*}
The continuity equation, via (\ref{b32}), yields
\begin{gather*}  \dfrac{\dot{a}}{a} +\dfrac{\dot{\alpha}}{\alpha} (m+1) +\dfrac{\dot{\beta}}{\beta}(m-1)  = 0 ,
\qquad \dfrac{\dot{c}}{c} + \dfrac{\dot{\alpha}}{\alpha} (m-1)+
\dfrac{\dot{\beta}}{\beta}(m+1) = 0,  %\label{5.7.33}
\end{gather*}
whence
\begin{gather} a = c_{\mathrm{I}}\alpha^{-(m+1)}\beta^{1-m} ,\qquad c = c_{\mathrm{II}}\alpha^{1-m}\beta^{-(m+1)} .\label{5.7.4}
\end{gather}

Moreover, (\ref{b33}) shows that
\begin{gather*}
\rho_0 =c_{\mathrm{III}}(\alpha\beta)^{1-m}=c^*_{\mathrm{III}}
\Omega^{2(1-m)} .%\label{5.7.5}
\end{gather*}
In the above, $c_{\mathrm{I}}$, $c_{\mathrm{II}}$, $c_{\mathrm{III}}$ and
$c^*_{\mathrm{III}}$ are arbitrary non-zero constants of
integration. The momentum equation gives
\begin{gather}
\ddot{\alpha}+2\varepsilon_2(t)\frac{m}{m-1}a\alpha=0 ,\qquad
\ddot{\beta}+2\varepsilon_2(t)\frac{m}{m-1}c\beta=0  \label{5.7.6}
\end{gather}
together with
\begin{gather*}
\ddot{\bar{p}}=0 ,\qquad \ddot{\bar{q}}=0  .%\label{5.7.7}
\end{gather*}

  Insertion of the expressions (\ref{5.7.4}) into
(\ref{5.7.6}) gives
\begin{gather*}
\ddot{\alpha}+2\varepsilon_2(t)\dfrac{m}{m-1}\dfrac{c_{\mathrm{I}}}{\alpha^2\beta}(\alpha\beta)^{2-m}=0 ,\qquad
\ddot{\beta}+2\varepsilon_2(t)\dfrac{m}{m-1}\dfrac{c_{\mathrm{II}}}{\alpha\beta^2}(\alpha\beta)^{2-m}=0
 ,%\label{5.7.62}
\end{gather*}
whence, in view of the relation (\ref{a55}), we again obtain a
Ermakov--Ray--Reid system, namely
\begin{gather}
\ddot{\alpha}=\frac{c^*_{\mathrm{I}}}{\alpha^2\beta} ,\qquad
\ddot{\beta}=\frac{c^*_{\mathrm{II}}}{\alpha\beta^2},\label{5.7.8}
\end{gather}
with the Ray--Reid invariant
\begin{gather*}
I=\frac{1}{2}(\dot{\alpha}\beta-\alpha\dot{\beta})^2+c^*_{\mathrm{I}}\frac{\beta}{\alpha
}+c^*_{\mathrm{II}}\frac{\alpha}{\beta}, %\label{5.7.9}
\end{gather*}
where
\begin{gather*}
c^*_{\mathrm{I}}=-2\alpha
c_{\mathrm{I}}\left(\frac{c^*_{\mathrm{III}}}{c_{\mathrm{III}}}\right)^{\frac{m-2}{m-1}} ,\qquad
c^*_{\mathrm{II}}=-2\alpha
c_{\mathrm{II}}\left(\frac{c^*_{\mathrm{III}}}{c_{\mathrm{III}}}\right)^{\frac{m-2}{m-1}} .%\label{5.7.10}
\end{gather*}
It is observed moreover, that the system (\ref{5.7.8}) is also
Hamiltonian with additional integral of motion
\begin{gather*}
H=\frac{1}{2}\big(c_{\mathrm{I}}\dot{\beta}^2+c_{\mathrm{II}}\dot{\alpha}^2\big)+\frac{c^*_{\mathrm{I}}
c^*_{\mathrm{II}}}{\alpha\beta} .%\label{5.7.11}
\end{gather*}

\section{A Lax pair formulation}

 It is now shown, following a procedure analogous to that set down in the spinning gas cloud analysis of~\cite{JPA12}, that the nonlinear dynamical system (\ref{b32}) and (\ref{b39})
admits an associated Lax pair representation.
In this connection, it is seen that the nonlinear dynamical system
given by~(\ref{b32}) together with~(\ref{b39}) arising from the
ansatz~(\ref{b30}) and~(\ref{b31}) may be written in the compact
matrix form as
\begin{gather*}   \dot{\mathbf{E}} + \mathbf{EL}+\mathbf{L}^T\mathbf{E}+(m-1)\mathbf{E} \operatorname{tr} \mathbf{L}=\textbf{0} ,\\
\dot{\mathbf{L}} +
\mathbf{L}^2+f  \mathbf{PL}+2\varepsilon_2(t)\dfrac{m}{m-1}\mathbf{E}=\textbf{0},
 \qquad m \neq 1, %\label{5.6.2}
\end{gather*}
where $\mathbf{L}$, $\mathbf{E}$ are given by (\ref{b31}) and
\begin{gather*}
 \mathbf{P}=  \begin{pmatrix} 0 & -1 \\ 1 & 0 \end{pmatrix}. %\label{5.6.3}
 \end{gather*}
Moreover, the relations (\ref{b33}) and (\ref{b40}) yield
\begin{gather*}
 \dot{\rho}_0+(m-1)\rho_0\operatorname{tr} \mathbf{L}=0 %\label{5.6.4}
 \qquad \mbox{and}\qquad
 \dot{\mathbf{M}}+f\mathbf{PM}=\textbf{0} . %\label{5.6.5}
\end{gather*}

  A gauge transformation is now introduced via
\begin{gather*}
\tilde{\mathbf{L}}=\mathbf{DLD}^{-1}+\frac{1}{2}f\mathbf{P} ,\qquad
\tilde{\mathbf{E}}=\mathbf{DED}^{-1},  %\label{5.6.6}
\end{gather*}
where
\begin{gather*}
\mathbf{D}=\exp\left(\frac{1}{2}\mathbf{P}ft\right)  %\label{5.6.7}
\end{gather*}
to obtain
\begin{gather} \dot{\tilde{\mathbf{E}}} + \tilde{\mathbf{E}}\tilde{\mathbf{L}}+\tilde{\mathbf{L}}^T
\tilde{\mathbf{E}}+(m-1)\tilde{\mathbf{E}} \operatorname{tr}
\tilde{\mathbf{L}}=\mathbf{0}\label{5.6.8}
\end{gather}
and
\begin{gather}
\dot{\tilde{\mathbf{L}}} +
\tilde{\mathbf{L}}^2+\dfrac{1}{4}f^2\mathbf{I}+2\varepsilon_2(t)\dfrac{m}{m-1}\tilde{\mathbf{E}}=\mathbf{0}.
 \label{5.6.9}
\end{gather}

  On use of the relation
\begin{gather*}
\mathbf{P}{\mathbf{H}}\mathbf{P}={\mathbf{H}}^T-(\operatorname{tr}
{\textbf{H}}) \textbf{I}  %\label{5.6.10}
\end{gather*}
 together with the Cayley--Hamilton identity
\begin{gather*}
\tilde{\mathbf{L}}^2-(\operatorname{tr}
\tilde{\textbf{L}})\tilde{\textbf{L}}+(\det
\tilde{\textbf{L}})\textbf{I}=\textbf{0}  %\label{5.6.11}
\end{gather*}
it is seen that (\ref{5.6.9}) yields
\begin{gather}
  \dot{\tilde{\mathbf{L}}} +(\operatorname{tr} \tilde{\mathbf{L}})\tilde{\mathbf{L}}-(\det \tilde{\mathbf{L}}) \mathbf{I}
 +\dfrac{1}{4} f^2\mathbf{I} + 2\varepsilon_2(t)\dfrac{m}{m-1}\tilde{\mathbf{E}}=\textbf{0} .
\label{5.6.12}
\end{gather}
Moreover, on introduction of a new trace-free matrix
$\tilde{\textbf{Q}}$ according to
\begin{gather*}
\tilde{\textbf{Q}}=\textbf{P}\tilde{\mathbf{E}} %\label{5.6.13}
\end{gather*}
the matrix equation (\ref{5.6.8}) becomes
\begin{gather}
\dot{\tilde{\mathbf{Q}}} +[\tilde{\mathbf{Q}}, \tilde{\mathbf{L}}]+
m \tilde{\mathbf{Q}}(\operatorname{tr} \tilde{\mathbf{L}})=0 .\label{5.6.14}
\end{gather}

  Since $\operatorname{tr}\textbf{L}=\operatorname{tr}
\tilde{\textbf{L}}=2\dot{\Omega}/\Omega$, it is natural to introduce
the scaling
\begin{gather*}
\bar{\textbf{L}}=\tilde{\mathbf{L}}\Omega^2 ,\qquad \bar{\textbf{E}}=\tilde{\mathbf{E}}\Omega^{2m} ,\qquad
\bar{\textbf{Q}}=\tilde{\mathbf{Q}}\Omega^{2m},%\label{5.6.15}
\end{gather*}
whence (\ref{5.6.12}) and (\ref{5.6.14}) reduce, in turn, to
\begin{gather}
\dot{\bar{\mathbf{L}}}-\Omega^{-2}(\det \bar{\mathbf{L}})\mathbf{I}+\dfrac{f^2}{4}\Omega^{2}\mathbf{I}
+2\varepsilon_2(t)\dfrac{m}{m-1} \Omega^{2(1-m)}\bar{\mathbf{E}}=\textbf{0}
 \label{5.6.16}
\end{gather}
and
\begin{gather}
\dot{\bar{\mathbf{Q}}}+\Omega^{-2} [\bar{\mathbf{Q}}, \bar{\mathbf{L}}]=\textbf{0},
\label{5.6.17}
\end{gather}
where
\begin{gather*}
\bar{\mathbf{L}}^*=\bar{\mathbf{L}}-\frac{1}{2}(\operatorname{tr} \bar{\mathbf{L}})\mathbf{I} %\label{5.6.18}
\end{gather*}
denotes the trace-free part of $\bar{\mathbf{L}}$. Moreover, the
trace-free part of (\ref{5.6.16}) yields
\begin{gather}
\dot{\bar{\mathbf{L}}}^*+\varepsilon_2(t)\dfrac{m}{m-1} \Omega^{2(1-m)}[\bar{\mathbf{Q}},
\mathbf{P}]=\textbf{0},
  \label{5.6.19}
\end{gather}
while its trace gives
\begin{gather}
(\operatorname{tr}\bar{\mathbf{L}})^\cdot-2\Omega^{-2}(\det \bar{\mathbf{L}}^*)-\dfrac{1}{2}\Omega^{-2}(\operatorname{tr}\bar{\mathbf{L}})^2
+\dfrac{1}{2}f^2\Omega^2+2\varepsilon_2(t)\dfrac{m}{m-1} \Omega^{2(1-m)}(\operatorname{tr} \bar{\mathbf{E}})=0 .
 \label{5.6.20}
\end{gather}

  Insertion of the expression (\ref{a55}) for
$\varepsilon_2$ (with $\alpha = 1$) into (\ref{5.6.19}) yields
\begin{gather}\label{5.6.21}
\dot{\bar{\mathbf{L}}}^*+\Omega^{-2}[\bar{\mathbf{Q}}, \mathbf{P}]=\textbf{0}
\end{gather}
and on introduction of the new time measure $\tau$ according to
\begin{gather*}
d\tau=\Omega^{-2}d t  %\label{5.6.22}
\end{gather*}
the systems (\ref{5.6.17}) and (\ref{5.6.21}) become in turn
\begin{gather}
\bar{\mathbf{Q}}'+[\bar{\mathbf{Q}}, \bar{\mathbf{L}}^*]=\textbf{0}
 %\label{5.6.23}
\qquad \mbox{and} \qquad
\label{5.6.24}
\bar{\mathbf{L}}^{*\prime}+[\bar{\mathbf{Q}},\mathbf{P}]=\textbf{0},
\end{gather}
where the prime denotes $d/d \tau$. The matrix system
(\ref{5.6.24}) constitutes the compatibility
condition
\begin{gather*}
\mathcal {M'(\lambda)}+[\mathcal {M(\lambda)} , \mathcal
{L(\lambda)}] =\textbf{0}%\label{5.6.25}
\end{gather*}
for the Lax pair
\begin{gather}
\Psi'=\mathcal {L(\lambda)}\Psi  ,\qquad \mu\Psi=\mathcal
{M(\lambda)}\Psi, \label{5.6.26}
\end{gather}
where
\begin{gather*}
\mathcal {L(\lambda)}=\bar {\mathbf{L}}^*+\lambda \mathbf{P}
 ,\qquad \mathcal {M(\lambda)}=\bar {\mathbf{Q}}+\lambda \bar
{\mathbf{L}}^*+\lambda^2\mathbf{P} .%\label{5.6.27}
\end{gather*}
An analogous result has been obtained in the case of non-conducting
rotating gas clouds in~\cite{RS11}. As in that work, there is an
interesting Steen--Ermakov connection. Thus, on setting
\begin{gather*}
\Sigma=\Omega^{-1}
\end{gather*}
then the relation (\ref{5.6.20}) is readily shown to reduce to a
Steen--Ermakov equation, namely
\begin{gather*}
\Sigma ''+(\det \bar{\mathbf{L}}^*-\operatorname{tr} \bar{\mathbf{E}})\Sigma=\dfrac{f^2}{4\Sigma^3} .
%\label{5.6.30}
\end{gather*}
Results of \cite{RS11} related to the Lax pair for a spinning gas cloud system carry over \textit{mutatis mutandis} to the Lax pair~(\ref{5.6.26}) obtained in the present magnetogasdynamic study. Thus, the linear system~(\ref{5.6.26}) is gauge equivalent to the standard Lax pair for the stationary reduction of the integrable cubic nonlinear Schr\"{o}dinger equation. The connection may be made in the manner set down in~\cite{RS11}.

\section{Conclusion}

It has been shown via an elliptic vortex ansatz that there is hidden
integrable structure of Ermakov--Ray--Reid type underlying a
2+1-dimensional non-isothermal magnetogasdynamic system. The Ermakov
variables turn out to have a natural physical interpretation as the
semi-axes of the time-modulated density representation. Moreover, a
Lax pair for the original non\-li\-near dynamical subsystem has been
constructed. The preceeding and previous studies such as that in
\cite{EFKK04} suggest that a general investigation of the occurrence
of integrable Ermakov--Ray--Reid structure in 2+1-dimensional
hydrodynamic systems would be of interest. It is noted that Hamiltonian--Ermakov type systems have been additionally investigated in \cite{CL91,HG96}.

\pdfbookmark[1]{References}{ref}
\LastPageEnding

\end{document}